\documentclass[aps,prl,twocolumn,showpacs,superscriptaddress,floatfix]{revtex4}
\usepackage{amsfonts}
\usepackage{amsmath}
\usepackage{amssymb}
\usepackage{bm}
\usepackage{graphics,graphicx}
\usepackage{psfrag}
\usepackage{epsfig}
\usepackage{bbm}

\def\pra{ Phys. Rev. A}

\def\etal{{\it et al.\/}}

\newcommand{\be}{\begin{equation}}
\newcommand{\ee}{\end{equation}}
\newcommand{\bea}{\begin{eqnarray}}
\newcommand{\eea}{\end{eqnarray}}

\newcommand{\nn}{\nonumber}

\newcommand{\up}{\uparrow}
\newcommand{\down}{\downarrow}

\newcommand{\ket}[1]{\left\vert #1    \right\rangle }
\newcommand{\bra}[1]{\left\langle   #1  \right\vert}

\begin{document}

\title{Fast cooling of trapped ions using the dynamical Stark shift gate}

\author{A. Retzker} \email{a.retzker@imperial.ac.uk}
\author{M.B. Plenio}

\affiliation{Institute for Mathematical Sciences, Imperial College
London, SW7 2PE, UK} \affiliation{QOLS, The Blackett Laboratory,
Imperial College London, Prince Consort Rd., SW7 2BW, UK}

\date{\today}

\begin{abstract}
A laser cooling scheme for trapped ions is presented which is
based on the fast dynamical Stark shift gate, described in
[Jonathan \etal, \pra {\bf 62}, 042307]. Since this cooling method
does not contain an off resonant carrier transition, low final
temperatures are achieved even in traveling wave light field. The
proposed method may operate in either pulsed or continuous mode
and is also suitable for ion traps using microwave addressing in
strong magnetic field gradients.
\end{abstract}

 \maketitle

{\em Introduction --}
The ability to laser cool ions to within the vicinity of the
motional ground state is a key factor in the
realization of efficient quantum computation \cite{wineland1998num}.
Various cooling schemes have been suggested and implemented,
achieving lower and lower temperature and increased cooling
rates. The schemes range from the simplest Doppler cooling
\cite{Wineland1975,Hansch1975} to more sophisticated ion trap cooling methods,
including sideband cooling for two level atoms
\cite{diedrich1989,Wineland1975,Wineland1978}, Raman side band
cooling \cite{monroe1995} which may be used for ions and for
atoms\cite{hamann1998,vladan1998} and cooling schemes based on
electromagnetically induced transparency (EIT) \cite{morigi2000}.
The existing schemes may be divided into pulsed schemes
\cite{monroe1995} and continuous schemes.

The two-level and Raman sideband cooling schemes employ Rabi
frequencies that are typically an order of magnitude lower than the
trap frequency. This is necessary to avoid off-resonant excitations
which increase the average phonons number $\langle n \rangle$ and
thus lead to heating. Hence, these off-resonant excitations increase
the final temperature of the scheme. To overcome this limit, a
scheme that employs EIT was introduced in \cite{morigi2000} and
demonstrated in \cite{Roos2000}. This scheme uses interference to cancel
the undesirable off-resonant carrier transition by suppression of absorption.
This scheme can achieve low temperatures as well as
high cooling rates, but the exact calibration of the laser intensity and the
high intensities which are needed, makes the scheme challenging.

In this work we present an alternative approach to this problem
that relies on the fast Stark shift gate
\cite{Jonathan2000,Jonathan2001}. In this scheme off-resonant
carrier transitions that would lead to heating are forbidden even
when operating in the traveling wave regime at Rabi frequencies
that are of the order of the trap frequency. The higher
Rabi-frequency and the cancelation of the carrier transition
promises increased cooling rates and lower final temperatures.
This advantage applies both for the pulsed scheme
and the continuous cooling schemes.
It is important to note that unlike the original Stark shift gate
which requires careful intensity stabilization to achieve high
fidelity \cite{Jonathan2000,Jonathan2001} the Stark shift cooling
method is only weakly affected by laser intensity
fluctuations. Our scheme is applicable for trapped ions or atoms
and is also suitable for an ion trap scheme
employing microwave addressing method in strong magnetic field
gradients as proposed in \cite{Mintert2001}. The presented scheme
satisfies the dark state condition\cite{dum1994}, i.e. the
ions are pumped into a zero-phonon state that no longer interact
with the the applied laser fields.

This paper is organized as follows, We first describe the operation
of the Stark shift gate and then explain the idea of using this gate
for cooling both in pulsed and continuous operation. To
analyze our proposed scheme we deduce the rate equation in the
Lamb-Dicke limit and verify the predictions thus obtained with exact
numerical simulations. In order to discuss the efficiency of the
method we compare the proposed cooling scheme with previously
proposed cooling schemes. In order to demonstrate the applicability
of the scheme to chains of ions we present numerical results for the simultaneous
cooling of three ions. We finish with a discussion and
conclusion.

{\em The Stark-shift gate --} In the following we briefly explain
the basic mechanism underlying the Stark shift gate
\cite{Jonathan2000,Jonathan2001}. The Hamilton operator of a single
ion driven by a traveling-wave laser field of Rabi frequency
$\Omega$ in the standard interaction picture with respect to the free atom
and phonons is given by
\begin{displaymath}
    H= \Omega \{\sigma_+\exp(i\eta [a e^{-i\nu t }+a^\dagger e^{i\nu t } ]
    -i\delta t)+H.c\},
\end{displaymath}
where $\hbar=1$, $\delta$ is the laser atom detuning, $\nu$ the
trap frequency and $\eta$ is the Lamb-Dicke parameter
\cite{Wineland1998}. For $\delta=0$ and employing $\eta\ll 1$, the
exponent may be expanded to first order in $\eta$ yielding $H\cong
\Omega (\sigma_x +\eta \sigma_y (a e^{-i \nu t}+a^{\dagger} e^{i
\nu t}))$. In a further interaction picture with respect to $H_0=
\Omega \sigma_x$, we obtain $H_I = i \eta \Omega'
[e^{i(2\Omega-\nu)t}\sigma_+a
+e^{i(2\Omega+
\nu)t}\sigma_+a^\dagger
]+H.c.$ where $\sigma_+=|+\rangle\langle -|$. Setting
$\Omega=\frac{\nu}{2}$ (see also \cite{Lizuain06,Jonathan2000,Jonathan2001}) and neglecting
off resonant terms we obtain
\be H_{ss}=\frac{i \eta \nu}{2}[\sigma_+ a- \sigma_-
a^\dagger].\ee
Careful numerical investigations demonstrate that $H_{ss}$
represents an excellent approximation to the exact dynamics
\cite{Jonathan2000,Jonathan2001}. It is of particular importance
to note that the Stark-shift gate does not have off-resonant
carrier transitions, despite being operated in the traveling wave
regime, making it a very accurate and fast quantum gate. As the carrier transition represents
the dominant heating term in laser cooling, this cancellation of
the carrier transition suggests the use of the Stark shift gate
for fast laser cooling. The Stark shift gate requires
stabilization of the Rabi frequency at $\Omega=\frac{\nu}{2}$.
However, in laser cooling we are only interested in the
preparation of a target state -- the ground state. This suggests
that the constraints on the stability of the Rabi frequency are
much less severe in the application of the Stark shift gate to
laser cooling than for its gate operation. The verification and
exploitation of this idea is the purpose of the remainder of this
work.

{\em Stark-shift gate cooling --} In the following we will
describe how the Stark shift gate may be used to implement both
pulsed and continuous laser cooling schemes. We begin with the
pulsed scheme as this clarifies the origin of the speed advantage
in both schemes. We proceed in close analogy to the pulsed scheme
based on the regular gate \cite{monroe1995}. Assume that the
system starts in state $|\down\rangle|n\rangle$. First a resonant
laser pulse implements $|\down\rangle|n\rangle\rightarrow
|-\rangle|n\rangle$. Then the Stark shift gate creates the
rotation $\ket - \ket {n} \rightarrow \ket + \ket {n-1}$. A
resonant pulse then implements $\ket + \ket {n-1} \rightarrow \ket
\up \ket {n-1}$ hence moving the system to a dissipative level.
Spontaneous decay predominantly leads to $\ket \up \ket {n-1}
\rightarrow \ket \down \ket {n-1}$ and the process as a whole
leads to the net loss of one phonon. This set of pulses is
repeated until the limiting temperature is achieved.
This scheme yields lower final temperatures and is faster under
ideal conditions than the usual sideband cooling. It is evident
that the bottleneck of the scheme is the $\ket - \ket {n}
\rightarrow \ket + \ket {n-1}$ and it is this process that is
accelerated by the use of the Stark shift gate. The above scheme
may also be implemented in the microwave regime \cite{Mintert2001}
where instead of the cooling laser, a combination of a microwave
and a steep magnetic field gradient is used and the coupling to
the dissipative level is achieved by resonant lasers.

\begin{figure}
\begin{center}
\psfrag{e}{$\ket e$}
\psfrag{o}{$\Omega$}\psfrag{oc}{$\Omega_c$}\psfrag{g1}{$\ket
{g_2}$}\psfrag{g2}{$\ket {g_1}$} \psfrag{-0}{$\ket
{-,0}$}\psfrag{-1}{$\ket {-,1}$}\psfrag{-2}{$\ket {-,2}$}
\psfrag{+0}{$\ket {+,0}$}\psfrag{+1}{$\ket {+,1}$}\psfrag{+2}{$\ket
{+,2}$} \psfrag{e0}{$\ket {e,0}$}\psfrag{e1}{$\ket
{e,1}$}\psfrag{e2}{$\ket {e,2}$}
\includegraphics[width=0.40\textwidth,height=0.20\textheight]{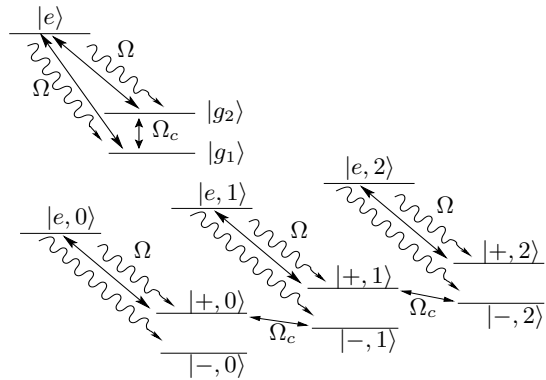}
\end{center}
\vspace*{-0.5cm} \caption{The required atomic level structure is
given in the left upper corner. A cooling Laser $\Omega_c$
manifests the Stark shift gate, creating a rotation $\ket -
\ket{n}\leftrightarrow \ket + \ket{n-1}$. The $\Omega_1$ and
$\Omega_2$ Lasers couple only the $\ket +$ state to the
dissipative level and leave the $\ket -$ state decoupled. The
cooling process is schematically depicted in the lower right part
of the picture. }\label{threelevel}
\end{figure}

The continuous scheme is most efficiently implemented in a three
level configuration as shown in Fig. \ref{threelevel}. The
Hamiltonian of this system is given by
\begin{eqnarray}
    H&=&\omega_1 \ket {g_1} \bra {g_1}+\omega_2 \ket {g_2}\bra
    {g_2} + \omega_3 \ket e \bra e +\nu a^\dagger a
    \nn\\
    &-& \Omega\left( \ket {g_1}  \bra {e} e^{-i(\omega_1+\Delta)
    t} + \ket {g_2} \bra {e} e^{-i(\omega_2+\Delta)t}+H.c\right) \nn\\
    &-&\Omega_{c}\left(\ket {g_1} \bra {g_2} e^{i(k_{c}x-\omega_{c}t)}
    + H.c\right). \label{Hamilton}
\end{eqnarray}
The laser on the meta-stable $|g_1\rangle \leftrightarrow
|g_2\rangle$ transition is used to implement the Stark shift gate
by tuning its Rabi frequency to the resonance condition
$\Omega_c=\frac{\nu}{2}$. This implements the rotation $\ket -
\ket {n} \rightarrow \ket + \ket {n-1}$. The two lasers on the
$|g_i\rangle\leftrightarrow |e\rangle$ transition are chosen to
have equal Rabi frequency and detuning and thus couple the $\ket
+=\frac{1}{\sqrt{2}}(|g_1\rangle+|g_2\rangle)$ state to the
dissipative level $\ket e$ with spontaneous decay rates
$2\Gamma_1$ and $2\Gamma_2$ to the states $\ket{g_1}$,$\ket{g_2}$
respectively. The two lasers leave the state $\ket -$ decoupled
from the dissipative level. It is important to note that the
lasers on the $|g_i\rangle\leftrightarrow |e\rangle$ do not couple
the internal and the phonon degrees of freedom, i.e, their
Lamb-Dicke parameter should be made as small as possible by
adjusting the angle between the laser and the ion trap.

The basic cooling cycle in the continuous scheme works as follows.
The laser with Rabi frequency $\Omega_c$ creates the rotation $\ket
- \ket n \leftrightarrow \ket+ \ket {n-1}$ based on the Stark shift
gate. The lasers couple the $\ket +\ket{n-1}$ state to the
dissipative level $|e\rangle\ket{n-1}$, which dissipates back to one of
the $\ket {g_i,n-1}$'s which are
superpositions of $\ket +\ket{n-1}$ and $\ket -\ket{n-1}$. If the
decay is to the $\ket -\ket{n-1}$, one phonon has been lost and
cooling is achieved, whereas if the decay is to the $\ket
+\ket{n-1}$ state the cycle is repeated. Thus on average phonons
are dissipated out of the system. Note that this mechanism is
independent of the sign of the detuning $\Delta$.

\begin{figure}
\psfrag{W}{W}
\psfrag{e}{$\eta$}
\begin{center}
\includegraphics[width=0.45\textwidth,height=0.20\textheight]{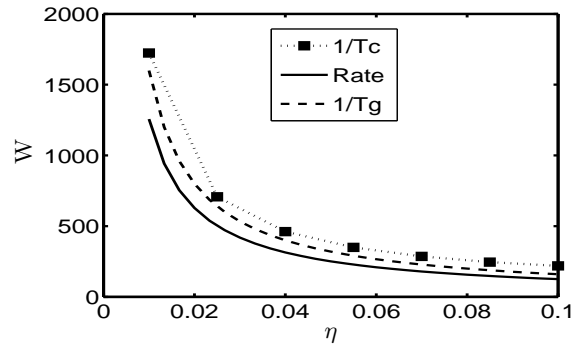}
\end{center}
\vspace*{-0.5cm} \caption{The Rate at the optimal point. The squares
are $\frac{1}{T_{C}}$, when $T_{C}$ is the time that takes to reduce
the population from $\langle n \rangle=1$ to $\langle n
\rangle=0.01$. The blue line corresponds to half a period of a Rabi
frequency and the green line to the cooling rate calculated from the
rate equations $W= \frac{1}{8} \eta \Omega_c $ }\label{TimeRate}
\end{figure}

\begin{figure}
\psfrag{o}{$\Omega_c$}
\psfrag{n}{$\langle n \rangle$}
\begin{center}
\includegraphics[width=0.45\textwidth,height=0.20\textheight]{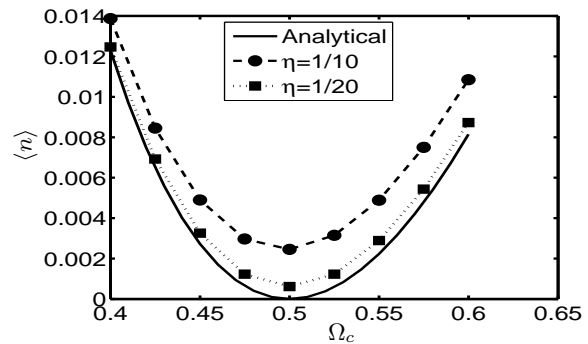}
\end{center}
\vspace*{-0.5cm} \caption{$\langle n \rangle$ as a function of
$\Omega_c$. The minimum is at the Stark shift resonance. The
continuous line is the rate equation result. The squares are
calculated from a numerical solution of the Master equation
for $\eta=1/20$ and the triangles are for
$\eta=1/10$. In the limit $\eta \rightarrow 0$ The numerics
coincide with the rate equations. The parameters are
$\Gamma=10,\Omega=1/10,\nu=1$ and $\Delta=0$ }\label{ro}
\end{figure}

{\em Analytical Results --} Following the heuristic arguments
presented so far we will now derive analytic expressions for the
cooling rate and final temperature of our scheme based on a Master
equation approach. The dynamics and the final state can be
obtained from the master equation,
\be \frac{\partial}{\partial t}\rho=\frac{1}{i}[H,\rho]+\mathcal{
L}\rho, \label{master}\ee
where $H$ is the Hamiltonian eq. (\ref{Hamilton}) and $\mathcal{L}$
is the Liouville operator describing the dissipative channels of
from level $|e\rangle$ to $|g_i\rangle$. In the following we use the
techniques presented in \cite{lindberg1986,cirac1992}. In this
formalism, the equation of motion for the phonons is obtained by
adiabatic elimination of the internal degrees of freedom of the
ions. The final temperature and the cooling rate are obtained by
expanding the Liouville operator and the density matrix in eq.
(\ref{master}) in the Lamb-Dicke parameter
$\mathcal{L}=\mathcal{L}_0+\eta \mathcal{L}_1+\eta^2
\mathcal{L}_2+\mathcal{O}(\eta^3)$, $\rho=\rho_0+\eta \rho_1+\eta^2
\rho_2+\mathcal{O}(\eta^3)$. Inserting this into eq. (\ref{master})
results in three equations, each for a different power of the
Lamb-Dicke parameter as in \cite{lindberg1986}. From these three
equations the final temperature and the cooling rate are calculated.
The validity of the expansion in the Lamb-Dicke parameter at the
point $\Omega_c=\frac{\nu}{2}$ is restricted by the three conditions
$\Gamma\nu\eta,\nu^2\eta, \Delta\nu\eta \ll \Omega^2$.
The result of this approach is a rate equation for the phonon
probability distribution $P(n)$ \cite{Javanainen1984}
\bea
    \frac{d}{dt}P(n)&=& \eta^2 \Big(A_-[(n+1)P(n+1)-n P(n)] \nn\\
&& \hspace*{0.6cm} + A_+ [n P(n-1)-(n+1) P(n)]\Big)
\eea
where
\be A_-(\nu) \!=\! \frac{2 \Gamma \Omega ^2 \Omega_c^2}{\Gamma ^2
(\nu -2 \Omega_c)^2+\left(2 \Omega ^2+(\nu -2 \Omega_c) (\Delta
-\nu
   +\Omega_c)\right)^2}
\ee
and $A_+=A_-(-\nu)$. From these expressions the final population
and the rate are deduced to be
\bea \langle n\rangle \!\!\!&=&\!\!
\frac{A_+}{A_--A_+}\label{meanphonon}\\
&&\hspace*{-0.5cm} =\!\frac{\Gamma ^2 (\nu -2 \Omega_c)^2+\left(2
\Omega ^2+(\nu -2 \Omega_c) (\Delta -\nu +\Omega_c)\right)^2}{4
\nu \left(2 \Omega_c \Gamma ^2+(\Delta +3 \Omega_c) \left(\nu ^2+2
\left(\Omega_c (\Delta +\Omega_c)-\Omega
^2\right)\right)\right)}\nn\\
W \!\!\!&=&\!\! \eta^2 (A_--A_+),\eea where $W$ is the cooling
rate.
%
Numerics shown in Fig. \ref{ro}  confirms that the  $\langle n
\rangle$ is minimized at the Stark shift gate point $\Omega_c
\approx \frac{\nu}{2}$. At $\Omega_c = \frac{\nu}{2}$ we find $W =
\frac{1}{8} \Gamma \eta ^2 \nu ^2 \Omega ^2 (\Omega ^{-4}-4[4 \Gamma
^2 \nu ^2+(3 \nu ^2+2 \Delta \nu -2 \Omega ^2)^2]^{-1} )$. The
choice $\frac{\Gamma\nu}{\Omega^2}\eta = 1$ realizes a cooling
scheme with a rate $W = \frac{1}{8}\eta \Omega_c$ which is close to
the inverse of the gate time $T=\frac{\pi}{\eta \Omega_c}$. Note
that this choice saturates the conditions of validity in the above
derivation, If one of the other conditions is saturated as well the
final population is smaller by the factor of 2, which means that the
choice $\Delta=\Gamma$ is optimal. Nevertheless, numerical results
shown in Fig. \ref{TimeRate} corroborate that the cooling time is of
the order of the gate time.
Fig. \ref{ro} shows numerically and analytically that for the
final $\langle n\rangle$ as a function of $\Omega_c$ we obtain a
parabola with a minimum around the Stark shift resonance. At this
point we find \be\langle n \rangle=\frac{\Omega ^4}{\nu \left(\nu
\Gamma ^2+\left(\Delta +\frac{3 \nu }{2}\right) \left(\nu ^2+2
\left(\frac{1}{2} \left(\Delta +\frac{\nu }{2}\right) \nu -\Omega
^2\right)\right)\right).}\ee Inserting the values of the optimum
point, $\Gamma\nu\eta,\nu^2\eta, \Delta\nu\eta = \Omega^2$, yields
a temperature proportional to $\eta^2$. For $\Omega=\nu/2$ heating
$A_+$ is weaker than cooling $A_-$ whenever $\Omega^2\le\frac{\nu
\left(4 \Gamma ^2+4 \Delta ^2+9 \nu ^2+12 \Delta \nu \right)}{4 (2
\Delta +3 \nu )}$. The cooling efficiency is not strongly effected
by the fluctuations of the laser. For example,
at the optimal point, at $\eta=0.05$, for intensity fluctuations of
10 percent, the rate and the final temperature change by less then 50 percent.

\begin{figure}
\psfrag{na}{$\langle n \rangle$}
\begin{center}
\includegraphics[width=0.45\textwidth,height=0.20\textheight]{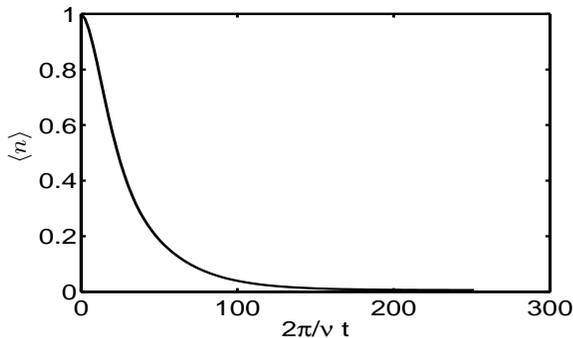}
\end{center}
\vspace*{-0.5cm} \caption{$\langle n\rangle$ as a function of time
using Monte Carlo simulation. The parameters used for the
simulation:
$\Delta=0$,$\Gamma=6$,$\nu=1$,$\Omega_c=1/2$,$\Omega_1=\Omega_2=1$,$\eta=1/10$.
For these parameters the gate time is $\frac{\pi}{\eta \Omega}=
62.5$. It could be seen from the numerics that the cooling time is
in the same order of magnitude as the gate time.}\label{cooling1}
\end{figure}

{\em Numerical results --} The results deduced from the rate
equations are correct in the limit $\eta \rightarrow 0$. In order to
check the results for finite $\eta$ and see the dynamics, we compare
these results to a numerical solution of the Master equation. In Fig.
\ref{ro} a plot of the analytical final mean phonon number is compared to
numerical results and as $\eta\rightarrow 0$ the
rate equation results are reproduced. In order to compare the rates
found for the final stages of cooling with the actual cooling time
we reproduce the dynamics using Monte - Carlo simulation for the
cooling of one phonon. Fig. \ref{cooling1} shows the dynamics of the
mean phonon number.

The mean phonon number in sideband cooling is $\langle
n\rangle=(\alpha+\frac{1}{4})\left(\frac{\Gamma}{\nu}\right)^2$
,where $\alpha$ is a geometric factor of the order $1$
\cite{Stenholm1986,Javanainen1981c}. In order for sideband cooling to
operate at the same rate as the proposed cooling, $\Gamma$ has to be
of the order of $ \nu/2 \eta$ which means that the sidebands are not
resolved, hence, sideband cooling could not be applied. The cooling
rates and temperatures found in our scheme could be achieved by the
EIT cooling method \cite{morigi2000} but in this method the laser
power should be at least an order of magnitude stronger then in the
cooling described in this work.

\begin{figure}[h]
\includegraphics[width=0.45\textwidth,height=0.25\textheight]{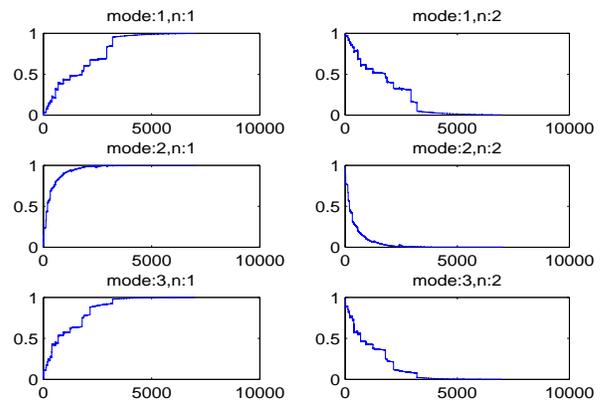}
\vspace*{-0.8cm} \caption{Monte-Carlo simulation for the cooling of three ions. The Rabi frequency
is set to $\Omega_c=\frac{1}{2}\nu_2+0.15$. For three ions the
frequencies are $\nu_2=\sqrt{3}\nu_1,\nu_3=\sqrt{29/5}\nu_1$.}
\label{chain}
\end{figure}

{\em Multi-mode cooling --} Due to the cancelation of the carrier
transition it suggests itself that our cooling method based on the
Stark shift gate could be used to cool several modes simultaneously.
Fig.\ref{chain} shows Monte-Carlo simulation \cite{Plenio1998} for the
cooling of three modes in a three ion chain. Here we chose the laser Rabi frequency near the
second mode frequency. The cooling rate achieved in this way of course
slower than for a single ion, but still faster than side band cooling.
In a different scheme individual ions could be addressed with lasers
of different Rabi frequency each at the optimal working point for a
different mode. Thus the whole chain could be cooled at times of the
order the Stark shift gate time.

{\em Conclusions --} We have introduced a method for
the fast cooling of an ion trap which operates with rates of the
fast Stark shift gate\cite{Jonathan2000}. The final temperatures
achieved are proportional to $\eta^2$(recoil energy). The cooling method is
efficient both in continuous and pulsed operation,
achieving lower temperatures and faster rates then ordinary
sideband cooling. The gate is also suitable for ion traps with
microwave frequency addressing \cite{Mintert2001}.

{\em Acknowledgements --} We would like to thank H. Haeffner, N.
Davidson and B. Reznik for helpful discussions. This work has been
supported by the European Commission under the Integrated Project
Qubit Applications (QAP) funded by the IST directorate as Contract
Number 015848, the Royal Society and is part of the EPSRC QIP-IRC.

\end{document}